\documentclass{article}
\usepackage{graphicx}
\usepackage{amsmath}

\input{tcilatex}

\begin{document}

\begin{titlepage}

\begin{center}

\vspace{1cm}

{\Large \textbf{Differences between scalar field and scalar density solutions%
}}

\vspace{1cm}

{\bf Nurettin Pirin{\c c}{\c c}io{\~g}lu$%
^{a}$ and Ilker Sert$^{b}$}

\vspace{0.5cm}

{\it $^{a}$ Department of Physics, University of Dicle, Diyarbak\i r, TR21280.\\[%
0pt] $^{b}$ Department of Physics, Institute of Nuclear Science,
Izmir, TR35100.}

\end{center}

\vspace{1cm}

\begin{abstract}
We explore differences between scalar field, and scalar density
solutions by using Robertson-Walker (RW) metric, and also a
non-relativistic Hamiltonian is derived for a scalar density field
in the post-Newtonian approximation. The results are compared with
those of scalar field. The expanding universe in RW metric, and
post-Newtonian solution of Klein-Gordon equation are separately
discussed.
\end{abstract}

\end{titlepage}

\section{Introduction}

There is a vital role of tensors in physics, but the role of the
tensor densities in physics is not discussed in literatures. Just
we know properties of it, but there is no application to
understand the effects it may cause, except that using
$\sqrt{\left( -g\right) }$ to make the volume element, $d^{4}x$, a
tensor in general relativity. So let us remember some properties
of scalar densities; for example, the determinant of a metric
tensor is a tensor density \cite {Weinberg}

\begin{equation}
g\equiv \det g_{\mu \nu }.
\end{equation}

The transformation rule for the metric tensor is

\begin{equation}
g_{\mu ^{\prime }\nu ^{\prime }}(x^{\prime })=\frac{\partial x^{\sigma }}{%
\partial x^{\mu ^{\prime }}}\frac{\partial x^{\rho }}{\partial x^{\nu
^{\prime }}}g_{\sigma \rho }(x),
\end{equation}

and taking determinants of the transformation, we have

\begin{equation}
\left( -g^{\prime }\right) =J^{2}\left( -g\right)
\end{equation}

Hence the metric determinant $g$ is a scalar density of weight $+2$ \cite
{D'Inverno}. Taking the square roots, to get

\begin{equation}
\sqrt{\left( -g^{\prime }\right) }=J\sqrt{\left( -g\right) },
\end{equation}

and so the $\sqrt{-g}$ is a scalar density of weight $+1.$  Beside the covariant derivative of $g_{\mu \nu }$%
, the covariant derivative of $\sqrt{-g}$ is trivial, that is

\begin{equation}
\nabla _{\mu }\sqrt{-g}\equiv 0.
\end{equation}

Any tensor density of weight $W$ can be expressed as an ordinary tensor times a factor $%
g^{-W/2}$ \cite{Weinberg}. Thus, if we have a scalar density,
$\Omega $ of weight $+1$, and we can construct a scalar identity
dividing $\Omega $ by $\sqrt{-g}.$ Thus, we have an ordinary
scalar,

\begin{equation}
\left( \frac{\Omega }{\sqrt{-g}}\right)
\end{equation}

where both $\sqrt{-g}$, and $\Omega $ are scalar densities of
weight $+1$. The covariant differentiation of this quantity is not
different from that of the ordinary scalar field. That is,

\begin{equation}
\nabla _{\mu }\left( \frac{\Omega }{\sqrt{-g}}\right) \equiv \partial _{\mu
}\left( \frac{\Omega }{\sqrt{-g}}\right) .
\end{equation}

Our idea is that the quantity $\left( \frac{\Omega
}{\sqrt{-g}}\right) ,$ can be used to solve many problems in
physics and we expect the solutions will be different when
compared with scalar field solutions.Some properties of scalar
field, and scalar density of weight $+1$ are summarized in
\textbf{Table 1}.

\vspace{0.5cm}

\textbf{Table 1}. Properties of scalar field and scalar density
field.\\
\begin{tabular}{lll}

\hline
\textbf{Scalar field, }$\phi $ &  & \textbf{Scalar density, }$\Omega $ \\
\hline
& \textbf{Transformation} &  \\
\begin{tabular}{l}
$x_{\mu }\longrightarrow $ $x_{\mu }^{^{\prime }}$, \\
$\phi (x)\longrightarrow \phi (x^{^{\prime }})$%
\end{tabular}
&  &
\begin{tabular}{l}
$x_{\mu }\longrightarrow $ $x_{\mu }^{^{\prime }},$ \\
$\Omega (x)\longrightarrow \det \left( \frac{\partial x^{^{\prime }}}{%
\partial x}\right) \Omega (x^{^{\prime }})$%
\end{tabular}
\\
&
\begin{tabular}{l}
\textbf{\textbf{Covariant}} \\
\textbf{\textbf{Derivative}}
\end{tabular}
&  \\
$\nabla _{\mu }\phi =\partial _{\mu }\phi $ &  & $\nabla _{\mu }\Omega
=\partial _{\mu }\Omega -\Gamma _{\alpha \mu }^{\alpha }\Omega $ \\
& \textbf{Kinetic Terms} &  \\
\begin{tabular}{l}
minimal kinetic term: \\
$K=\frac{1}{2}g^{\mu \nu }\partial _{\mu }\phi \partial _{\nu }\phi $%
\end{tabular}
&  &
\begin{tabular}{l}
non-minimal kinetic term: \\
$K=\frac{1}{2}g^{\mu \nu }\left[ \partial _{\mu }\Omega -\Gamma _{\alpha \mu
}^{\alpha }\Omega \right] $ \\
$\ \ \ \ \ \ \ \ \ \ \times \left[ \partial _{\nu }\Omega -\Gamma _{\beta
\nu }^{\beta }\Omega \right] $%
\end{tabular}
\\
& \textbf{Potential} &  \\
$
\begin{tabular}{l}
depends only on $\phi .$ \\
Potential does not \\
contribute to evolution \\
of $\phi $'s equation and \\
gravity if $V(\phi )=0.$%
\end{tabular}
$ &  &
\begin{tabular}{l}
depends both on $\Omega $, and \\
$\sqrt{-g}.$ Potential does not \\
contribute to gravity, but \\
$\Omega $'s evolution is affected if \\
$V\left( \frac{\Omega }{\sqrt{-g}}\right) =M_{1}^{3}\frac{\Omega }{\sqrt{-g}}%
.$%
\end{tabular}
\\ \hline
\end{tabular}

The parameter $M_{1}$ in the \textbf{Table 1} is the mass
parameter of scalar field.

\vspace{0.5cm}

By using the scalar density the expanding universe in the RW
metric, and the post-Newtonian solution of Klein-Gordon equation
are separately discussed. The paper is organized as follows;
section 2 is the calculations for scalar field and scalar density
field in RW metric, section 3 is a derivation of Hamiltonian in
post-Newtonian approximation, and ended with a short conclusion.

\section{Equation of motion for a scalar, and a scalar density field in the
RW metric}

Recent observations such as luminosity distance-redshift relation for
supernova Ia \cite{Riess et al} \cite{Perlmutter98}, gravitational lansing
\cite{Mellier99}, velocity fields \cite{Juszkiewicz}, and the cosmic
microwave background temperature anisotropies \cite{Melchiorri} suggest that
the universe is currently dominated by an energy component (matter) with a
negative pressure \cite{Wang} \cite{Krauss95} \cite{Perlmutter99} \cite
{Schmidt}. The possible candidates for that energy part is cosmological
constant, dynamical vacuum energy or a homogeneous scalar field
(quintessence) that is very weakly coupled to ordinary matter \cite{Peebles
and Ratra} \cite{Peebles} \cite{Ratra} \cite{Chiba2001}.

An action integral for a homogeneous scalar field theory is given
as

\begin{equation}
S\left[ \phi \right] =\int d^{4}x\sqrt{-g}\left[ -\frac{1}{2}M_{pl}^{2}R+%
\frac{1}{2}g^{\mu \nu }\nabla _{\mu }\phi \nabla _{\nu }\phi -M_{1}^{3}\phi
-V(\phi )\right] .  \label{action1}
\end{equation}

Where $M_{pl}$ is the Planck mass ($M_{pl}^{2}=\frac{1}{8\pi G}$),
$R$ is the Ricci curvature of spacetime, $\nabla _{\mu }$ is the
covariant derivative operator. On very large scales the universe
is spatially homogeneous and isotropic, which implies that its
metric takes the RW form \cite{Carroll2000}

\begin{equation}
ds^{2}=dt^{2}-a^{2}(t)\left[ \frac{dr^{2}}{1-\kappa r^{2}}+r^{2}d\Omega ^{2}%
\right]
\end{equation}

Where $d\Omega ^{2}=d\theta ^{2}+\sin ^{2}\theta d\varphi ^{2},$ $a(t)$ is
the cosmological expansion scale factor, and curvature parameter $\kappa $
takes values $+1$, $0$, and $-1$ for positively, flat, and negatively curved
spaces, respectively. Varying equation (\ref{action1}) with respect to $\phi
,$ and $g_{\mu \nu }$ to get , respectively, the Klein-Gordon equation, and
Friedmann equation:

\begin{equation}
\ddot{\phi}+3H\dot{\phi}+M_{1}^{3}+\frac{dV}{d\phi }=0,  \label{KG1}
\end{equation}

\begin{equation}
H^{2}=\frac{1}{M_{pl}^{2}}\left\{ \frac{1}{2}\dot{\phi}^{2}-\left[
M_{1}^{3}\phi +V(\phi )\right] \right\} .  \label{F1}
\end{equation}

Where dot represents the time derivative, the Hubble expansion
rate is defined by $H\equiv \frac{\dot{a}}{a}$. If we set the
$V=0,$ $M_{1}$ is still affects the evolution of $\phi .$ The
scalar field density, $\rho _{\phi }$, and the mass density
associated with it, $\rho _{M_{1}}$, drive the expansion. Where
$\rho _{\phi }\equiv{\dot{\phi}^{2}}$, and $\rho _{M_{1}}
\equiv{M_{1}^{3}\phi}$.

In the scalar density picture; the action (\ref{action1}) can be
written as follows

\begin{eqnarray}
S\left[ \Omega \right] &=&\int d^{4}x\sqrt{-g}[-\frac{1}{2}M_{pl}^{2}R+\frac{%
1}{2}g^{\mu \nu }\nabla _{\mu }\left( \frac{\Omega }{\sqrt{-g}}\right)
\nabla _{\nu }\left( \frac{\Omega }{\sqrt{-g}}\right)  \notag \\
&&-M_{1}^{3}\left( \frac{\Omega }{\sqrt{-g}}\right) -V\left( \frac{\Omega }{%
\sqrt{-g}}\right) ].  \label{action2}
\end{eqnarray}

Where $\Omega $ is a scalar density of weight $W=+1$, and the determinant of
metric, $g$, is a scalar density of weight $W=+2$ \cite{D'Inverno}.
Covariant derivative of a scalar density of weight $W$ is, $\nabla _{\mu
}\Omega =\partial _{\mu }\Omega -W\Gamma _{\alpha \mu }^{\alpha }\Omega $ .
Varying action (\ref{action2}) with respect to the line-element, one gets
the energy-momentum tensor as

\begin{eqnarray}
T_{\mu \nu } &=&\frac{1}{M_{pl}^{2}}\{-\frac{1}{2}g_{\mu \nu }g^{\alpha
\beta }\nabla _{\alpha }\left( \frac{\Omega }{\sqrt{-g}}\right) \nabla
_{\beta }\left( \frac{\Omega }{\sqrt{-g}}\right) \frac{\Omega }{\sqrt{-g}}
\notag \\
&&+g_{\mu \nu }[V\left( \frac{\Omega }{\sqrt{-g}}\right) -\left( \frac{%
\Omega }{\sqrt{-g}}\right) V^{^{\prime }}\left( \frac{\Omega }{\sqrt{-g}}%
\right) ]+5\nabla _{\mu }\left( \frac{\Omega }{\sqrt{-g}}\right) \nabla
_{\nu }\left( \frac{\Omega }{\sqrt{-g}}\right)   \notag \\
&&-g_{\mu \nu }g^{\alpha \beta }\nabla _{\alpha }\left( \frac{\Omega }{\sqrt{%
-g}}\nabla _{\beta }\left( \frac{\Omega }{\sqrt{-g}}\right) \right) \},
\label{T}
\end{eqnarray}

where prime denotes the derivative with respect to the metric.

Varying equation (\ref{action2}) with respect to $\Omega ,$ one
gets the field equation of scalar density $\Omega $ as

\begin{equation}
g^{\mu \nu }\nabla _{\mu }\nabla _{\nu }\left( \frac{\Omega }{\sqrt{-g}}%
\right) +M_{1}^{3}+\frac{dV\left( \frac{\Omega }{\sqrt{-g}}\right) }{d\Omega
}=0.  \label{KG}
\end{equation}

Contracting equation (\ref{T}) with $g_{\mu \nu }$ one gets the
following equation; by setting $V(\frac{\Omega }{\sqrt{-g}})=0$
and using the equation (\ref{KG});

\begin{equation}
R=\frac{1}{M_{pl}^{2}}\left\{ g^{\alpha \beta }\nabla _{\alpha }\left( \frac{%
\Omega }{\sqrt{-g}}\right) \nabla _{\beta }\left( \frac{\Omega }{\sqrt{-g}}%
\right) -4M_{1}^{3}\left( \frac{\Omega }{\sqrt{-g}}\right) \right\} .
\label{R}
\end{equation}

Consistent with the observations, cosmic microwave background (CMB) and Type
Ia supernova, the universe spatially homogeneous and isotropic, but evolving
in time \cite{Carroll}. In the RW metric the last two equations can be
written as follows;

\begin{equation}
\overset{\cdot \cdot }{\widetilde{\Omega }}-3H\overset{\cdot }{\widetilde{%
\Omega }}-3\overset{\cdot }{H}\widetilde{\Omega }+a^{3}(t)M_{1}^{3}=0,
\label{KG2}
\end{equation}

\begin{equation}
R=\frac{1}{M_{pl}^{2}}\left\{ \left( \frac{\overset{\cdot }{\widetilde{%
\Omega }}}{a^{3}}\right) ^{2}-6\frac{H}{a^{6}}\overset{\cdot }{\widetilde{%
\Omega }}\widetilde{\Omega }+\left( \frac{3H}{a^{3}}\right) ^{2}\widetilde{%
\Omega }^{2}-4\frac{M_{1}^{3}}{a^{3}}\widetilde{\Omega }\right\} .
\end{equation}

Here we used $\Omega =\frac{r^{2}\sin \theta }{\sqrt{1-\kappa r^{2}}}%
\widetilde{\Omega }$ transformation, and assumed that $\frac{\widetilde{%
\Omega }}{a^{3}}$ is completely homogeneous through space. Friedmann
equation for equation (\ref{action2}) is

\begin{equation}
H^{2}=\frac{1}{M_{pl}^{2}a^{6}}\left\{ \widetilde{\Omega }\left( \overset{%
\cdot \cdot }{\widetilde{\Omega }}-3\frac{\ddot{a}}{a}\widetilde{\Omega }%
\right) +\frac{3}{2}\left( \overset{\cdot }{\widetilde{\Omega }}\right) ^{2}+%
\frac{1}{2}H\widetilde{\Omega }\left( 11H\widetilde{\Omega }-8\overset{\cdot
}{\widetilde{\Omega }}\right) \right\}  \label{F2}
\end{equation}

The effect of $M_{1}^{3}$ is not seen in this equation, but by using
equation (\ref{KG2}), the last equation can be written as follows

\begin{equation}
H^{2}=\frac{3}{M_{pl}^{2}a^{6}}\left\{ \frac{1}{2}\left( \overset{\cdot }{%
\widetilde{\Omega }}\right) ^{2}+\frac{1}{3}\left( H\left( \frac{5}{2}H%
\widetilde{\Omega }-\overset{\cdot }{\widetilde{\Omega }}\right)
-a^{3}M_{1}^{3}\right) \widetilde{\Omega }\right\}  \label{F3}
\end{equation}

\begin{equation*}
\equiv \rho _{\Omega }+\rho _{M_{1}}
\end{equation*}

Comparing the solutions of scalar field in Eqs. (\ref{KG1}),
(\ref{F1}) with those of scalar density in Eq. (\ref{KG2}),
(\ref{F3}), they are completely different. Both the evolution
equation of scalar density, and Friedmann equation contain more
terms that drive the expansion when compared with those of scalar.
The third term in (\ref{F3}) is considered as a non-minimal coupling in ref.\cite{Chiba2001}. In equation (\ref{F3}), the density part of $%
\widetilde{\Omega }$ is proportional to $a^{-6}$, while that of
associated mass part is proportional to $a^{-3}$. So, the scalar
density part decreases with time more rapidly than the mass
density associated with it. In addition to the effect of $M_{1}$
the evolution of Eq. (\ref{KG2}) is affected by $a^{3}$.
\section{post-Newtonian approximation}

For a scalar field the Klein-Gordon equation can be written in the following
form;

\begin{equation}
g^{\mu \nu }\nabla _{\mu }\nabla _{\nu }\phi -\frac{m^{2}c^{2}}{\hbar }\phi
=0.  \label{KG0}
\end{equation}

The derivatives in the first term are covariant with respect to both general
coordinate and gauge transformation:

\begin{equation}
\nabla _{\mu }T^{\nu }\equiv \partial _{\mu }T^{\nu }+\Gamma _{\mu \sigma
}^{\nu }T^{\sigma }-\frac{ie}{\hbar c}A_{\mu }T^{\nu }
\end{equation}

for $T_{\mu }\equiv \partial _{\mu }\phi -\frac{ie}{\hbar c}A_{\mu }\phi .$
Where $\Gamma _{\mu \sigma }^{\nu }$ is the connection coefficients. In
post-Newtonian approximation $g^{\mu \nu }$ is

\begin{equation*}
g_{00}=-\left[ 1-2\frac{U}{c^{2}}+2\beta \left( \frac{U}{c^{2}}\right) ^{2}%
\right] ,
\end{equation*}

\begin{equation}
g_{ij}=\left[ 1+2\gamma \frac{U}{c^{2}}\right] \delta _{ij.}  \label{metric}
\end{equation}

where $U(x,t)$ is gravitational potential, $g_{0i}=0$, and $i$, $j=1$, $2$, $%
3$. Inserting (\ref{metric}) in (\ref{KG0}), and going to non-relativistic
limit , the equation of motion for (\ref{KG0}) can be written as follows\cite
{Laemmerzahl} \cite{0607113}:

\begin{equation}
i\hbar \partial _{t}\varphi =H\varphi
\end{equation}

where

\begin{eqnarray}
H_{\phi } &=&\frac{\vec{p}^{2}}{2m}-eA_{0}-mU-\left( \gamma +\frac{1}{2}%
\right) \frac{U}{mc^{2}}-\left( \frac{1}{2}-\beta \right) \frac{mU^{2}}{c^{2}%
}  \notag \\
&&-i\hbar \frac{\left( 2\gamma +1\right) }{2mc^{2}}\vec{g}.\vec{p}-\frac{%
3\gamma \pi G_{N}\hbar ^{2}}{mc^{2}}\rho _{m}.  \label{H_phi}
\end{eqnarray}

where $\rho _{m}$ is the rest mass density of the matter distribution, $\vec{%
g}=-\vec{\nabla}U$ being the gravitational acceleration. Now we consider the
Klein-Gordon equation for a scalar density,

\begin{equation}
g^{\mu \nu }\nabla _{\mu }\nabla _{\nu }\left( \frac{\Omega }{\sqrt{-g}}%
\right) -\frac{m^{2}c^{2}}{\hbar }\left( \frac{\Omega }{\sqrt{-g}}\right) =0,
\end{equation}

and using the same calculation techniques as done in scalar field in
post-Newtonian approximation, the Hamiltonian becomes

\begin{eqnarray}
H_{\Omega } &=&\frac{\vec{p}^{2}}{2m}-eA_{0}-mU-\left( \gamma +\frac{1}{2}%
\right) \frac{U}{mc^{2}}-\left( \frac{1}{2}-\beta \right) \frac{mU^{2}}{c^{2}%
}  \notag \\
&&-i\hbar \frac{\left( 8\gamma -3\right) }{2mc^{2}}\vec{g}.\vec{p}-\frac{%
(3\gamma -2)\pi G_{N}\hbar ^{2}}{mc^{2}}\rho _{m}.  \label{H_omega}
\end{eqnarray}

The Hamiltonian obtained for scalar density is different from that of scalar
field when comparing the coefficients of the last two terms.

\section{Conclusions}

The scalar density can be used for many problems in physics to
solve, and these solutions may rise interesting results, because
of the behavior of scalar density is different from that of scalar
itself. Obviously in the RW metric, the scalar density solution
has influenced both Klein-Gordon equation and Friedmann equation
when compared with those of scalar field. In the scalar density
solution; scalar density part decreases with time more rapidly
than the mass density associated with it. Also, in post-Newtonian
approximation it is clear to see that, scalar density solution
modifies both the Darwin term, and gravitation acceleration; the
effect of acceleration is increased $\simeq1.6$ times that of
scalar field solution in Einstein gravity.

\section*{Acknowledgements}

The authors would like to thank to D. A. Demir for the suggestion of the
research proposal and useful discussion.

\end{document}